\documentclass[11pt,pdftex]{article}

\usepackage{amsmath,amssymb}
\usepackage{graphicx}
\usepackage{moriond}
\usepackage[normalem]{ulem}

\newcommand{\dcp}{\delta_\textsc{cp}}
\newcommand{\Dmq}{\Delta m^2}
\newcommand{\eVq}{\ensuremath{\text{eV}^2}}
\newcommand{\LT}{\left}
\newcommand{\RT}{\right}

\newcommand{\EtAl}{\textit{et al.}}

\newcommand{\mysubsect}[1]{\par\medskip\noindent\textbf{\uline{#1}.}}

\newenvironment{myitemize} {
  \begin{list}{--}
      {
	\setlength{\leftmargin} {4mm}
	\setlength{\parsep}     {0pt}
	\setlength{\itemsep}    {0mm}
	\setlength{\topsep}     {\itemsep}
	\setlength{\partopsep}  {0pt}
	}} {
  \end{list}}

\begin{document}

\author{Michele Maltoni}

\address{Departamento de F\'isica Te\'orica \& Instituto de F\'isica Te\'orica,\\
  Facultad de Ciencias C-XI, Universidad Aut\'onoma de Madrid,\\
  Cantoblanco, E-28049 Madrid, Spain}

\title{POTENTIALITIES OF ATMOSPHERIC NEUTRINOS}

\maketitle

\abstracts{In this talk we will discuss the physics reach of the
  atmospheric neutrino data collected by a future megaton-class
  neutrino detector. After a general discussion of the potentialities
  of atmospheric neutrinos on general basis, we will consider concrete
  experimental setups and show that synergic effects exist between
  atmospheric and long-baseline neutrino data. Finally, we will show
  that present Super-Kamiokande data already have the capability to
  allow for a direct and unbiased measurement of the energy spectrum
  of the atmospheric neutrino fluxes.}

\section{Introduction}
\label{sec:introduction}

Despite their pioneering contribution to the discovery of neutrino
oscillations, it is in general assumed that atmospheric neutrino data
will no longer play an active role in neutrino physics in the coming
years. This is mainly due to the large theoretical uncertainties
arising from the poor knowledge of the atmospheric neutrino fluxes,
which strongly contrast with the requirement of ``precision'' needed
to further enhance our knowledge of the neutrino mass matrix. In this
talk, we will show that despite these large uncertainties the
atmospheric neutrino data collected by a megaton-class detector will
still provide very useful information.

Let us begin by reviewing what we have learned so far about neutrino
oscillations. The solar and atmospheric mass-squared differences are
clearly determined, and we know that the solar angle is large but
non-maximal while that the atmospheric angle is practically maximal.
The present best-fit point and $1\sigma$ ($3\sigma$) ranges
are:\,\cite{GonzalezGarcia:2007ib}
\begin{equation}\begin{aligned}
    \theta_{12} &= 33.7\pm 1.3\,\LT(_{-3.5}^{+4.3}\RT) \,,
    & \quad
    \;\Dmq_{21}\;
    &= 7.9\,_{-0.28}^{+0.27}\,\LT(_{-0.89}^{+1.1}\RT)
    \times 10^{-5}~\eVq \,,
    \\
    \theta_{23} &= 43.3\,_{-3.8}^{+4.3}\,\LT(_{-8.8}^{+9.8}\RT) \,,
    & \quad
    \LT|\Dmq_{31}\RT|
    &= 2.6 \pm 0.2\,(0.6) \times 10^{-3}~\eVq \,,
    \\
    \theta_{13} &= 0\,_{-0.0}^{+5.2}\,\LT(_{-0.0}^{+11.5}\RT) \,,
    & \quad
    \delta_\text{CP} &\in [0,\, 360] \,,
\end{aligned}\end{equation}
leading to the following values for the elements of the leptonic
mixing matrix, $U$, at 90\% CL:
\begin{equation}
    |U|_\text{90\%} =
    \begin{pmatrix}
	0.81 \to 0.85 & ~\quad~ 0.53 \to 0.58 & ~\quad~ 0.00 \to 0.12 \\
	0.32 \to 0.49 & ~\quad~ 0.52 \to 0.69 & ~\quad~ 0.60 \to 0.76 \\
	0.27 \to 0.46 & ~\quad~ 0.47 \to 0.64 & ~\quad~ 0.65 \to 0.80
    \end{pmatrix} ,
\end{equation}
and at the $3\sigma$ level:
\begin{equation}
    |U|_{3\sigma} =
    \begin{pmatrix} 
	0.79 \to 0.86 & ~\quad~ 0.50 \to 0.61 & ~\quad~ 0.00 \to 0.20 \\
	0.25 \to 0.53 & ~\quad~ 0.47 \to 0.73 & ~\quad~ 0.56 \to 0.79 \\
	0.21 \to 0.51 & ~\quad~ 0.42 \to 0.69 & ~\quad~ 0.61 \to 0.83
    \end{pmatrix} .
\end{equation}

\begin{figure}[t] \centering 
    \includegraphics[width=0.95\textwidth]{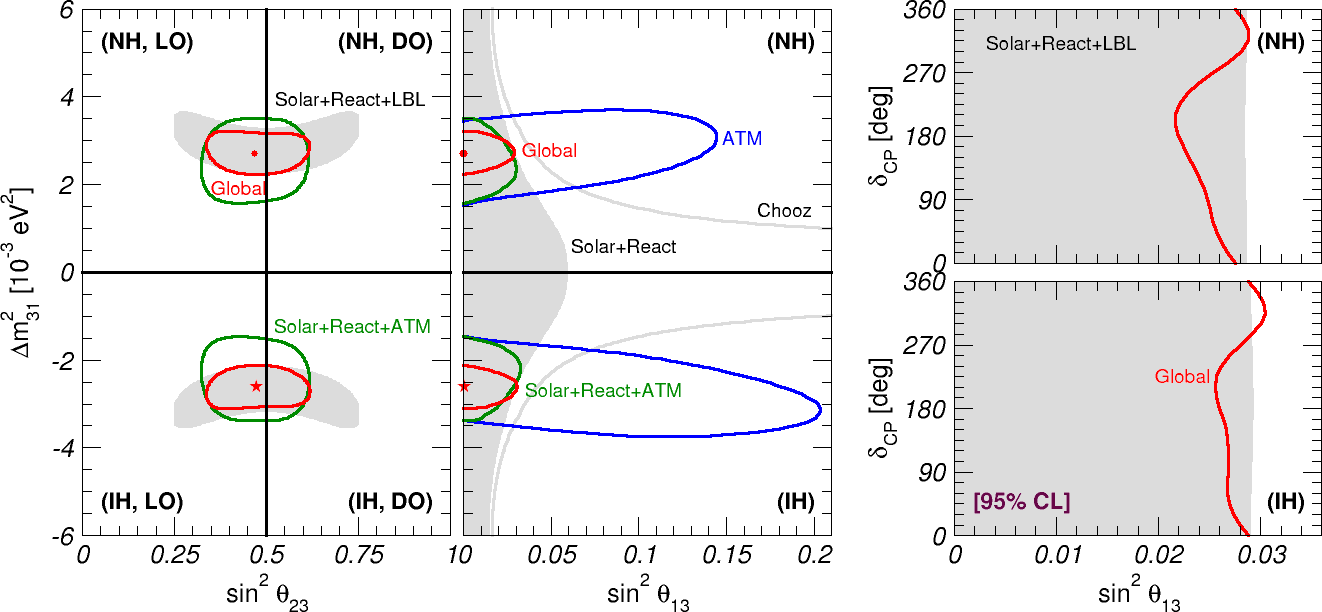}
    \caption{\label{fig:impact}%
      Allowed regions at 95\% CL from the analysis of different
      combinations of neutrino experiments. In particular, the gray
      regions correspond the combination of solar, reactor and
      accelerator data.\hfill~}
\end{figure}

To set the basis for the following discussion, it is useful to study
how much the information coming from atmospheric neutrino data
contributes to this picture. The ``solar'' parameters $\theta_{12}$
and $\Dmq_{21}$ are completely determined by the solar and KamLAND
data alone, and will therefore not be discussed here. As for the
remaining parameters, in Fig.~\ref{fig:impact} we show the allowed
regions implied by different combinations of neutrino experiments. It
is clear from this figure that the bound on $\theta_{13}$ comes mainly
from Chooz, with a small contribution from solar and KamLAND data,
while the atmospheric mass-squared difference $\Dmq_{31}$ is mainly
determined by the accelerator experiments K2K and Minos. The only
parameter whose determination is still dominated by atmospheric data
is $\theta_{23}$, being mainly a matter of total statistics. So it
seems that indeed atmospheric experiments already have not much to
say.

However, at a better look we note that present reactor and accelerator
data (gray regions) exhibit a very high degree of symmetry. In
particular, they have practically no dependence on $\dcp$, and they
are totally insensitive to the neutrino mass hierarchy (sign of
$\Dmq_{31}$) and to the octant (sign of $\theta_{23} - 45^\circ$).
Conversely, when atmospheric data are also included in the fit these
ambiguities, although far from being resolved, become at least
non-symmetric. In particular, the atmospheric bound on $\theta_{13}$
(blue lines) considerably depends on the mass hierarchy, and the
global fit with atmospheric data included (red lines) exhibit a weak
but visible dependence on the CP phase and on the octant.
Of course, this is not a proof that atmospheric data will be relevant
in the future, especially since the results of forthcoming accelerator
experiments such as T2K \emph{will} have a non-trivial dependence on
the value of $\dcp$ and on the neutrino mass hierarchy. However, it is
a hint the very broad-range information provided by atmospheric
neutrino data, which span about three orders of magnitude in length
and more than five in energy, may still be complementary to
accelerator experiments, which despite their high degree of precision
are limited to a fixed baseline and cover only a very limited range in
neutrino energy.
In the rest of this talk we will present a systematic study of these
potentialities.

\section{Sensitivity to oscillation parameters}
\label{sec:discussion}

As already mentioned, the strength of atmospheric data is its very
broad interval in neutrino energy ($E_\nu$) and baseline (determined
by the nadir angle $\Theta_\nu$). In order to provide a global view of
this whole range, in this section we will make extensive use of
neutrino oscillograms of the Earth, \textit{i.e.}\ contours of equal
probabilities in the neutrino energy--nadir angle
plane.\,\cite{Akhmedov:2006hb} More exactly, we will show contours of
equal oscillated-to-unoscillated ratio of atmospheric neutrino fluxes
of flavor $\beta$ arriving at the detector, $\Phi_\beta /
\Phi_\beta^0$, obtained by folding the primary neutrino flux
$\Phi_\alpha^0$ with the relevant conversion probability $P_{\alpha
\to \beta}$, so that $\Phi_\beta = \sum_\alpha \Phi_\alpha^0 \,
P_{\alpha \to \beta}$. As we will see, different regions of the
($E_\nu$, $\Theta_\nu$) plane will show characteristic structures
whose position and size is determined by various neutrino parameters.

\begin{figure}[t] \centering 
    \includegraphics[width=\textwidth]{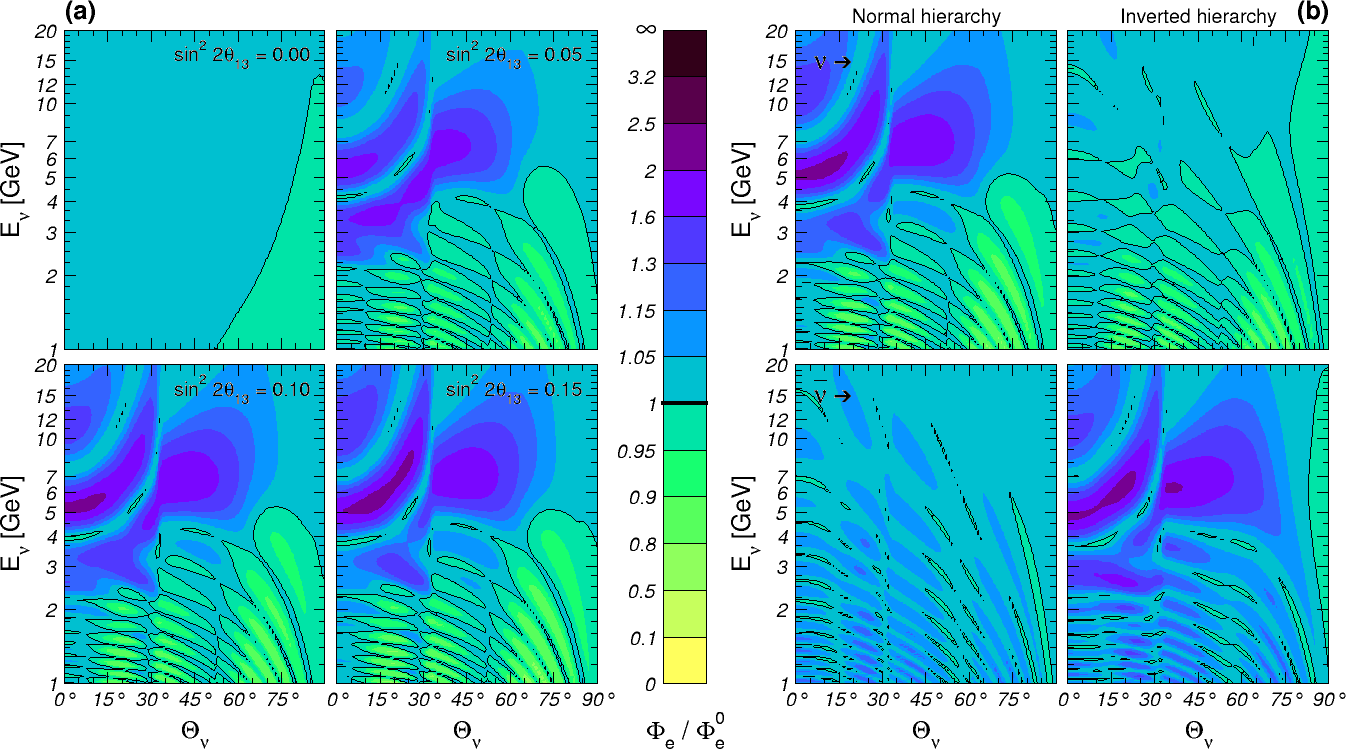}
    \caption{\label{fig:t13-hie}%
      Oscillograms of the oscillated-to-unoscillated flux ratio
      $\Phi_e / \Phi_e^0$. \uline{Left} (a): dependence on
      $\theta_{13}$. \uline{Right} (b): dependence on the neutrino
      mass hierarchy.\hfill~}
\end{figure}

\mysubsect{$\theta_{13}$}
Let us start by considering the sensitivity to $\theta_{13}$. As can
be seen from Fig.~\ref{fig:t13-hie}(a), for non-zero value of
$\theta_{13}$ matter effects induce a resonance in the $\nu_\mu \to
\nu_e$ conversion probability at $E_\nu \sim 3\div 6$~GeV. The precise
position of this peak in the $(E_\nu,\, \Theta_\nu)$ plane depends on
the value of $\theta_{13}$, so that in principle atmospheric neutrinos
could be used to measure this angle as long as it is larger than about
$3^\circ$. However, in practice the sensitivity is limited by two
factors:
\begin{myitemize}
  \item \emph{statistics}: at $E_\nu \sim 6$~GeV the atmospheric flux
    is already considerably suppressed;
    
  \item \emph{background}: the $\nu_\mu \to \nu_e$ signal is diluted
    by the unavoidable background of $\nu_e \to \nu_e$ events.
\end{myitemize}
Therefore, although some sensitivity is to be expected in a megaton
detector, it is likely that atmospheric neutrinos will not be
competitive with dedicated long-baseline and reactor experiments for
what concerns the determination of $\theta_{13}$. However, an explicit
observation of this resonance will provide a very important
confirmation of the MSW and parametric-resonance mechanisms.

\mysubsect{Hierarchy}
As shown in Fig.~\ref{fig:t13-hie}(b), the sensitivity to the
hierarchy arises from the observation of the same high-energy
resonance which is also involved in the sensitivity to $\theta_{13}$.
It is therefore only possible if $\theta_{13}$ is large enough. Note
that in order to determine the hierarchy it is not sufficient to see
the resonance (which would simply be a indication of non-zero
$\theta_{13}$), it is also necessary to tell whether it occurred for
neutrinos (normal hierarchy) or antineutrinos (inverted hierarchy). It
is therefore particularly important to have a detector capable of
charge discrimination. In the case of a non-magnetic detector such as
a Water Cerenkov, if a resonance is observed it might still be
possible to resolve the hierarchy by looking its \emph{size}, since
the number of neutrinos interacting in the detector is considerably
larger than the number of antineutrinos and therefore a normal
(inverted) hierarchy would result in a larger (smaller) signal. Note,
however, that the amplitude of the peak is affected both by
$\theta_{13}$ and by the value of $\theta_{23}$, as we will see in the
next paragraph, so that a poorly known value of these parameters will
result in a considerable loss of sensitivity.

\begin{figure}[t] \centering 
    \includegraphics[width=\textwidth]{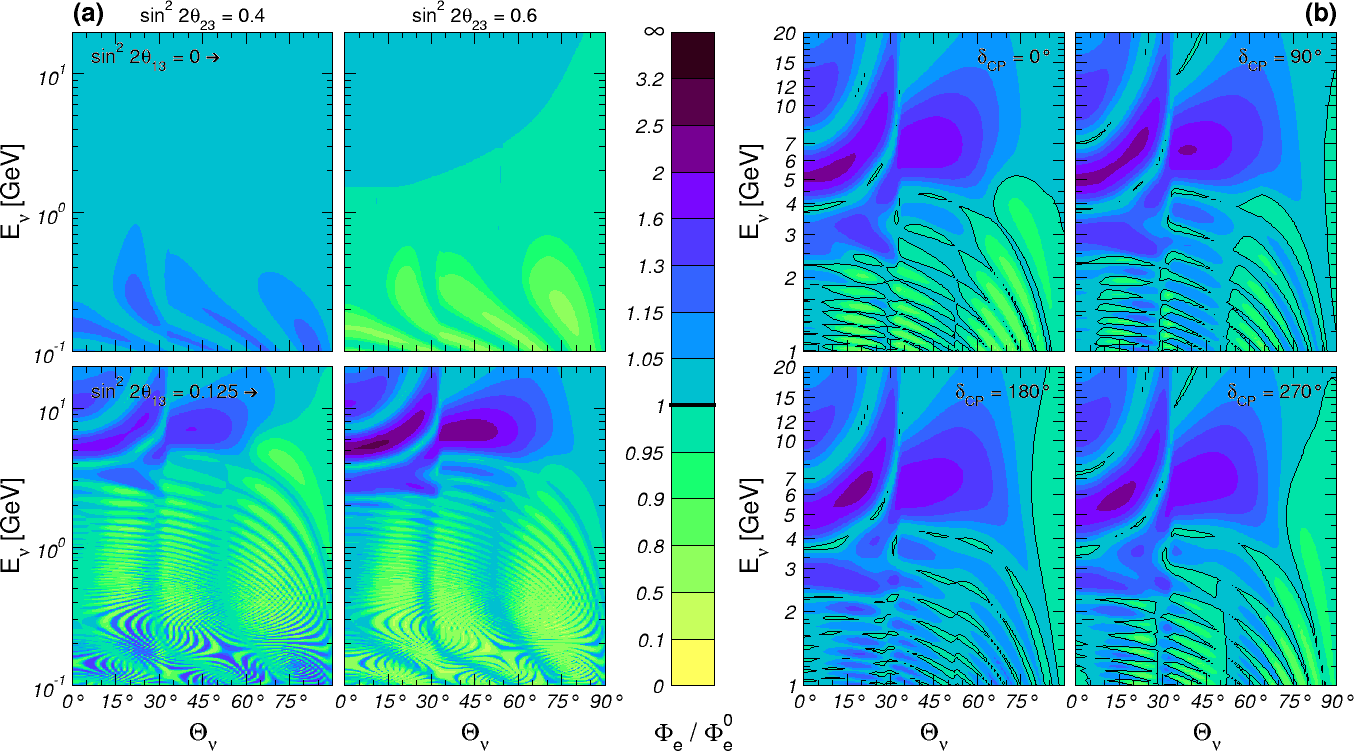}
    \caption{\label{fig:oct-pha}%
      Oscillograms of the oscillated-to-unoscillated flux ratio
      $\Phi_e / \Phi_e^0$. \uline{Left} (a): dependence on the
      $\theta_{23}$ octant, for $\theta_{13} = 0$ and $\sin^2
      2\theta_{13} = 0.125$. \uline{Right} (b): dependence on the CP
      phase for $\sin^2 2\theta_{13} = 0.125$.\hfill~}
\end{figure}

\mysubsect{Octant}
The sensitivity to the octant is one of the topics where atmospheric
neutrino are mostly useful. As can be seen Fig.~\ref{fig:oct-pha}(a),
we have two characteristic signatures:
\begin{myitemize}
  \item at \emph{low energy} ($E_\nu < 1$~GeV), we observe an excess
    (deficit) in the $\nu_e$ flux with respect to maximal mixing if
    $\theta_{23}$ is smaller (larger) than $45^\circ$. This effect is
    due to subleading oscillations induced by $\Dmq_{21}$, and is
    present also for $\theta_{13} = 0$. For $\theta_{13} \ne 0$ the
    $\nu_e$ flux arriving at the detector is modulated with the very
    fast $\Dmq_{31}$ oscillations, but the effect persists on average.
    Finally, this effect appears with the same sign for both neutrinos
    and antineutrinos, so that no charge discrimination is required
    for its identification.
    
  \item at \emph{high energy} ($E_\nu > 3$~GeV), we observe a decrease
    (increase) in the $\nu_e$ flux with respect to maximal mixing if
    $\theta_{23}$ is smaller (larger) than $45^\circ$. This effect is
    again related to the matter resonance discussed for $\theta_{13}$
    and the hierarchy, and indeed it appears only for $\theta_{13} \ne
    0$. As already seen, in a detector without charge discrimination
    the signal could be considerably suppressed.
\end{myitemize}
Note that the presence of a low-energy effect independent of
$\theta_{13}$ \emph{guarantees} a minimum sensitivity to the octant
from atmospheric neutrinos, provided that the deviation of
$\theta_{23}$ from maximal mixing is large enough. This is a unique
feature which will prove very synergic with long-baseline data, as we
will show in the next section. Note also that the slight preference
for $\theta_{23} < 45^\circ$ visible in Fig.~\ref{fig:impact} arises
precisely from this effect, and from the observation of a small excess
in sub-GeV $e$-like events in Super-Kamiokande data.

\mysubsect{CP phase}
Finally, let us spend a word on the sensitivity to the CP phase. A
characteristic signal is visible in the intermediate-energy region,
$1~\text{GeV} < E_\nu < 3$~GeV, and arises from the interference of
$\Dmq_{21}$-induced and $\Dmq_{31}$-induced oscillations. Although in
principle it is observable, as Fig.~\ref{fig:impact} demonstrates with
present data, this effect is quite small and probably hard to
disentangle from other parameters. Moreover, its presence depends
crucially on the size of $\theta_{13}$, so as for the sensitivity to
the hierarchy it is not ``guaranteed''. In the fits which we will
discuss in the rest of this talk the impact of $\dcp$ on the
determination of the other parameters is properly taken into account,
however we will not present any systematic study of the sensitivity of
atmospheric data to $\dcp$ itself.

\section{Synergies with long-baseline experiments}
\label{sec:results}

So far we have discussed the potentialities of atmospheric neutrinos
in general terms. Let us now consider concrete experimental setups and
compare their performances. In particular, we will focus on three
proposed experiments:\,\cite{Campagne:2006yx}
\begin{myitemize}
  \item a \uline{Beta Beam} ($\beta$B) from CERN to Fr\'ejus (130~Km).
    We assume 5 years of $\nu_e$ from $^{18}$Ne and 5 years of
    $\bar\nu_e$ from $^6$He at $\gamma = 100$, with an average
    neutrino energy $\LT< E_\nu \RT> = 400$~MeV. For the detector we
    assume the MEMPHYS Water-Cerenkov proposal, corresponding to 3
    tanks of 145~Kton each;
    
  \item a \uline{Super Beam} (SPL) from CERN to Fr\'ejus (130~Km).
    We assume 2 years of $\nu_\mu$ and 8 years of $\bar\nu_\mu$
    running, with an average energy $\LT< E_\nu \RT> = 300$~MeV. Again
    we use MEMPHYS as detector;
    
  \item the \uline{T2K phase II} (T2HK) experiment, corresponding to a
    4MW super beam from Tokai to Kamioka (295~Km), with 2 years of
    $\nu_\mu$ and 8 years of $\bar\nu_\mu$. The detector is the
    proposed Hyper-Kamiokande, rescaled to 440~Kton for a fair
    comparison with the $\beta$B and the SPL.
\end{myitemize}

\begin{figure}[t] \centering 
    \includegraphics[width=0.9\textwidth]{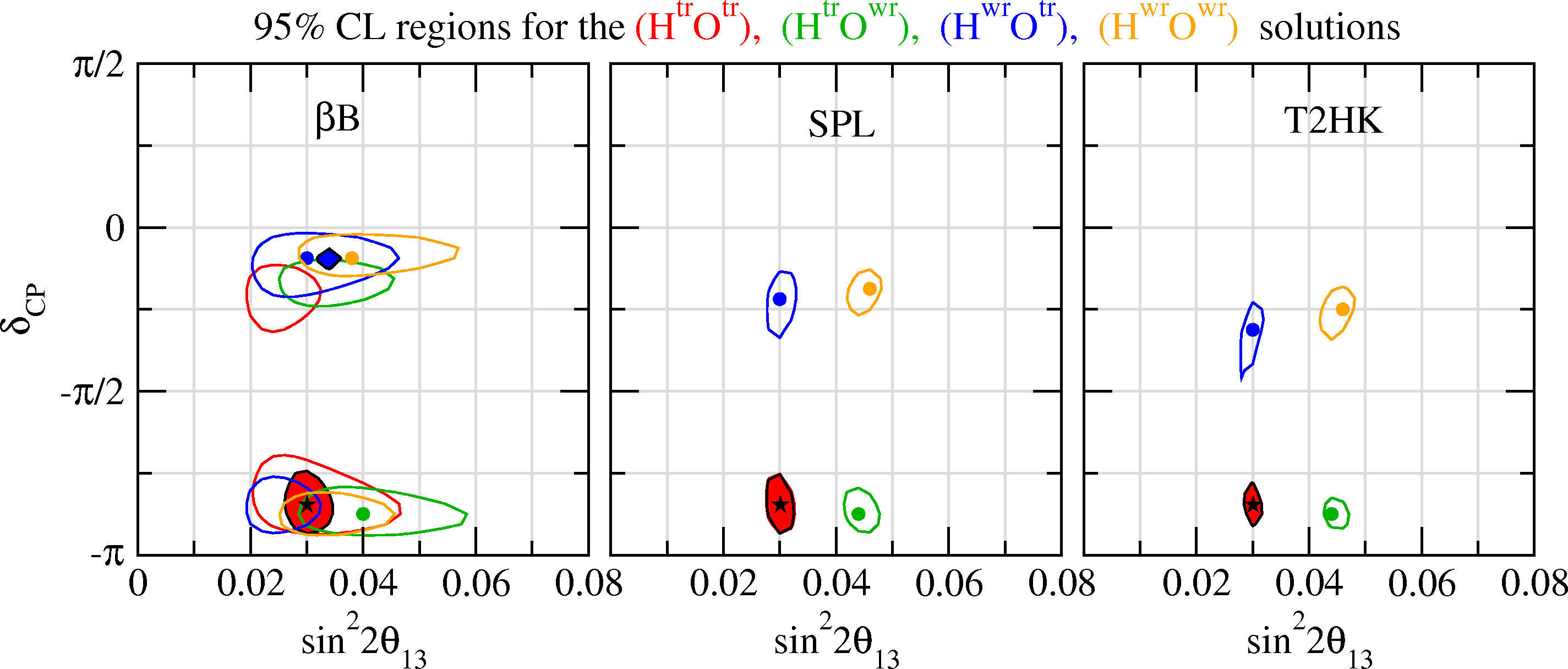}
    \caption{\label{fig:degeneracy}%
      Allowed regions in $\sin^22\theta_{13}$ and $\delta_\mathrm{CP}$
      for LBL data alone (contour lines) and LBL+ATM data combined
      (colored regions).\,\protect\cite{Campagne:2006yx}
      $\text{H}^\text{tr/wr}$ and $\text{O}^\text{tr/wr}$ refers to
      solutions with the true/wrong mass hierarchy and octant,
      respectively. The true parameter values are $\dcp = -0.85 \pi$,
      $\sin^2\theta_{12} = 0.3$, $\sin^22\theta_{13} = 0.03$,
      $\sin^2\theta_{23} = 0.6$, $\Dmq_{21} = 7.9\times 10^{-5}~\eVq$,
      $\Dmq_{31} = 2.4\times 10^{-3}~\eVq$.\hfill~}
\end{figure}

A characteristic feature in the analysis of future LBL experiments is
the presence of \emph{parameter degeneracies}.  Due to the inherent
three-flavor structure of the oscillation probabilities, for a given
experiment in general several disconnected regions in the
multi-dimensional space of oscillation parameters will be present.
Traditionally these degeneracies are referred to as follows:
\begin{myitemize}
  \item the \uline{intrinsic} degeneracy: for a measurement based on
    the $\nu_\mu \to \nu_e$ oscillation probability for neutrinos and
    antineutrinos two disconnected solutions appear in the
    ($\dcp,\,\theta_{13}$) plane;
    
  \item the \uline{hierarchy} degeneracy: the two solutions
    corresponding to the two signs of $\Dmq_{31}$ appear in
    general at different values of $\dcp$ and $\theta_{13}$;
    
  \item the \uline{octant} degeneracy: since LBL experiments are
    sensitive mainly to $\sin^22\theta_{23}$ it is difficult to
    distinguish the two octants $\theta_{23} < 45^\circ$ and
    $\theta_{23} > 45^\circ$.  Again, the solutions corresponding to
    $\theta_{23}$ and $\pi/2 - \theta_{23}$ appear in general at
    different values of $\dcp$ and $\theta_{13}$.
\end{myitemize}
This leads to an eight-fold ambiguity in $\theta_{13}$ and $\dcp$, and
hence degeneracies provide a serious limitation for the determination
of $\theta_{13}$, $\dcp$ and the sign of $\Dmq_{31}$.
In Fig.~\ref{fig:degeneracy} we illustrate this problem for the
$\beta$B, SPL and T2HK experiments. Assuming the true parameter values
$\delta_\mathrm{CP} = -0.85 \pi$, $\sin^22\theta_{13} = 0.03$ and
$\sin^2\theta_{23} = 0.6$ we show the allowed regions in the plane of
$\sin^22\theta_{13}$ and $\dcp$ taking into account the solutions with
the wrong hierarchy and the wrong octant.
As visible in this figure, for the $\beta$B the intrinsic degeneracy
cannot be resolved, due to the poor spectral information and the lack
of precise information on $|\Dmq_{31}|$ and $\sin^22\theta_{23}$
(usually provided by the $\nu_\mu$ disappearance), while for the super
beam experiments SPL and T2HK there is only a four-fold degeneracy
related to the sign of $\Dmq_{31}$ and the octant of $\theta_{23}$. On
the other hand, once atmospheric data are included in the fit all the
degeneracies are nearly completely resolved, and the true solution is
identified at 95\%~CL. This clearly show the presence of a synergy
between atmospheric and long-baseline data: at least for this specific
example, the combination of the two sets is much more powerful than
the simple sum of each individual data sample.

\begin{figure}[t] \centering 
    \includegraphics[height=53mm]{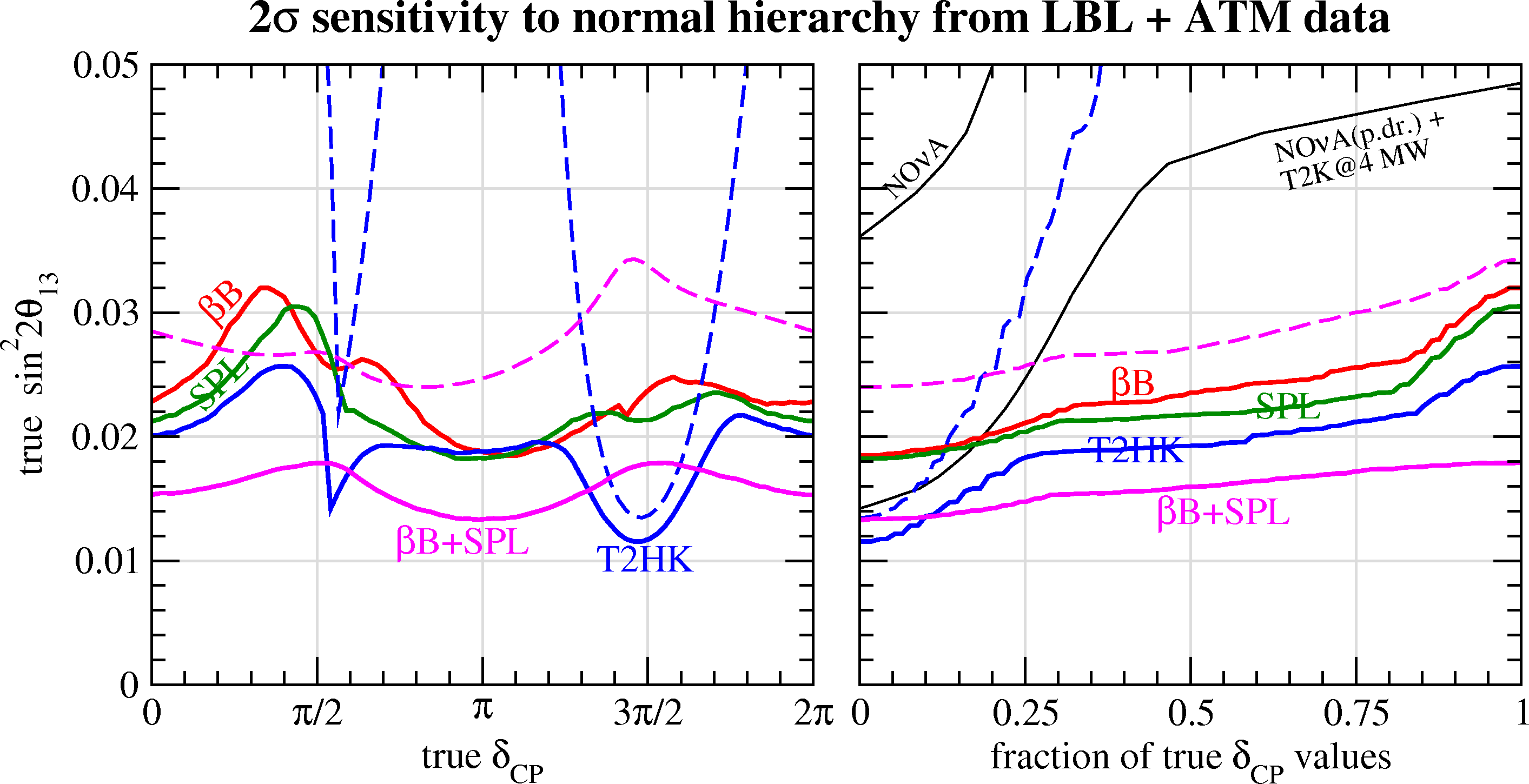}
    \hfil
    \raisebox{1pt}{\includegraphics[height=51mm]{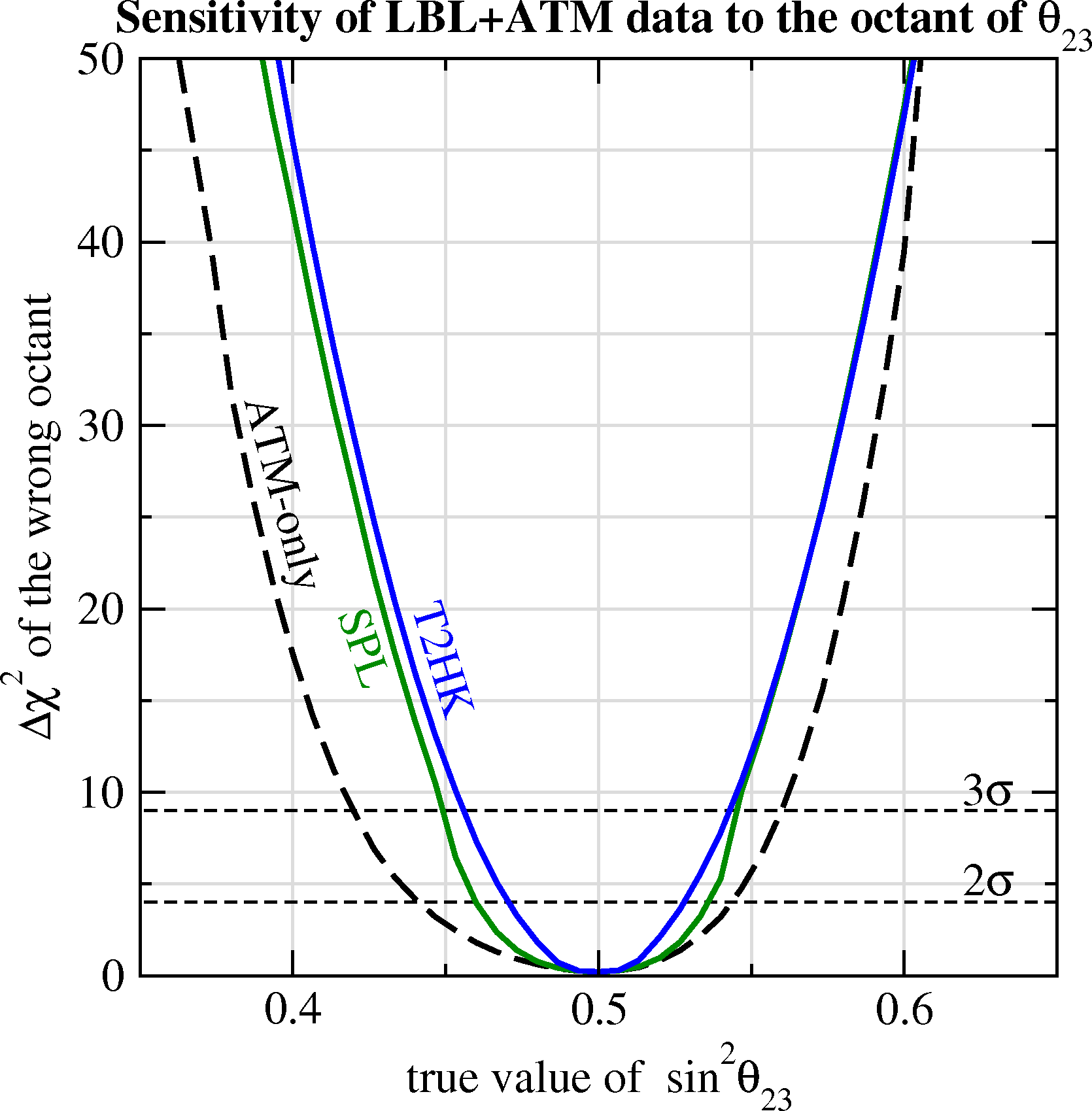}}
    \caption{\label{fig:resolve}%
      \uline{Left}: sensitivity to the mass hierarchy at $2\sigma$
      $(\Delta\chi^2 = 4)$ as a function of the true values of
      $\sin^22\theta_{13}$ and $\dcp$. The solid curves are the
      sensitivities from the combination of long-baseline and
      atmospheric neutrino data, the dashed curves correspond to
      long-baseline data only.
      \uline{Right}: $\Delta\chi^2$ of the solution with the wrong
      octant of $\theta_{23}$ as a function of the true value of
      $\sin^2\theta_{23}$. We have assumed a true value of
      $\theta_{13} = 0$.\hfill~}
\end{figure}

To further investigate this synergy, in the left panels of
Fig.~\ref{fig:resolve} we show how the combination of ATM+LBL data
leads to a non-trivial sensitivity to the neutrino mass hierarchy. For
LBL data alone (dashed curves) there is practically no sensitivity for
the CERN--MEMPHYS experiments (because of the very small matter
effects due to the relatively short baseline), and the sensitivity of
T2HK depends strongly on the true value of $\dcp$. However, by
including the data from atmospheric neutrinos (solid curves) the mass
hierarchy can be identified at $2\sigma$~CL provided
$\sin^22\theta_{13} \gtrsim 0.02 \div 0.03$. As an example we have
chosen in that figure a true value of $\theta_{23} = 45^\circ$; in
general the hierarchy sensitivity increases as $\theta_{23}$
increases.\,\cite{Huber:2005ep}
Note that the sensitivity to the neutrino mass hierarchy shown in
Fig.~\ref{fig:resolve} is significantly improved with respect to our
previous results.\,\cite{Huber:2005ep} There are two main reasons for
this better performance: first, we use now much more bins in charged
lepton energy for fully contained single-ring events; second, we
implemented also the information from multi-ring events. This latter
point is important since the relative contribution of neutrinos and
antineutrinos is different for single-ring and multi-ring events.
Therefore, combining both data sets allows to obtain a discrimination
between neutrino and antineutrino events on a statistical basis. This
in turn contains crucial information on the hierarchy, since as
discussed in Sec.~\ref{sec:discussion} the mass hierarchy determines
whether the matter enhancement occurs for neutrinos or for
antineutrinos.

In the right panel of Fig.~\ref{fig:resolve} we show the potential of
ATM+LBL data to exclude the octant degenerate solution. As seen in the
previous section, this effect is based mainly on oscillations with
$\Dmq_{21}$ and therefore we have very good sensitivity even for
$\theta_{13} = 0$; a finite value of $\theta_{13}$ in general improves
the sensitivity.\,\cite{Huber:2005ep} From the figure one can read off
that atmospheric data alone can resolve the correct octant at
$3\sigma$ if $|\sin^2\theta_{23} - 0.5| \gtrsim 0.085$. If atmospheric
data is combined with the LBL data from SPL or T2HK there is
sensitivity to the octant for $|\sin^2\theta_{23} - 0.5| \gtrsim
0.05$. The improvement of the octant sensitivity with respect to
previous analyses\,\cite{Huber:2005ep,GonzalezGarcia:2004cu} follows
from changes in the analysis of sub-GeV atmospheric events, where now
three bins in lepton momentum are used instead of one. Note that since
in this figure we have assumed a true value of $\theta_{13} = 0$,
combining the $\beta$B with ATM does not improve the sensitivity with
respect to atmospheric data alone.

\section{Direct determination of atmospheric fluxes}
\label{sec:fluxes}

\begin{figure}[t] \centering 
    \includegraphics[width=0.5\textwidth]{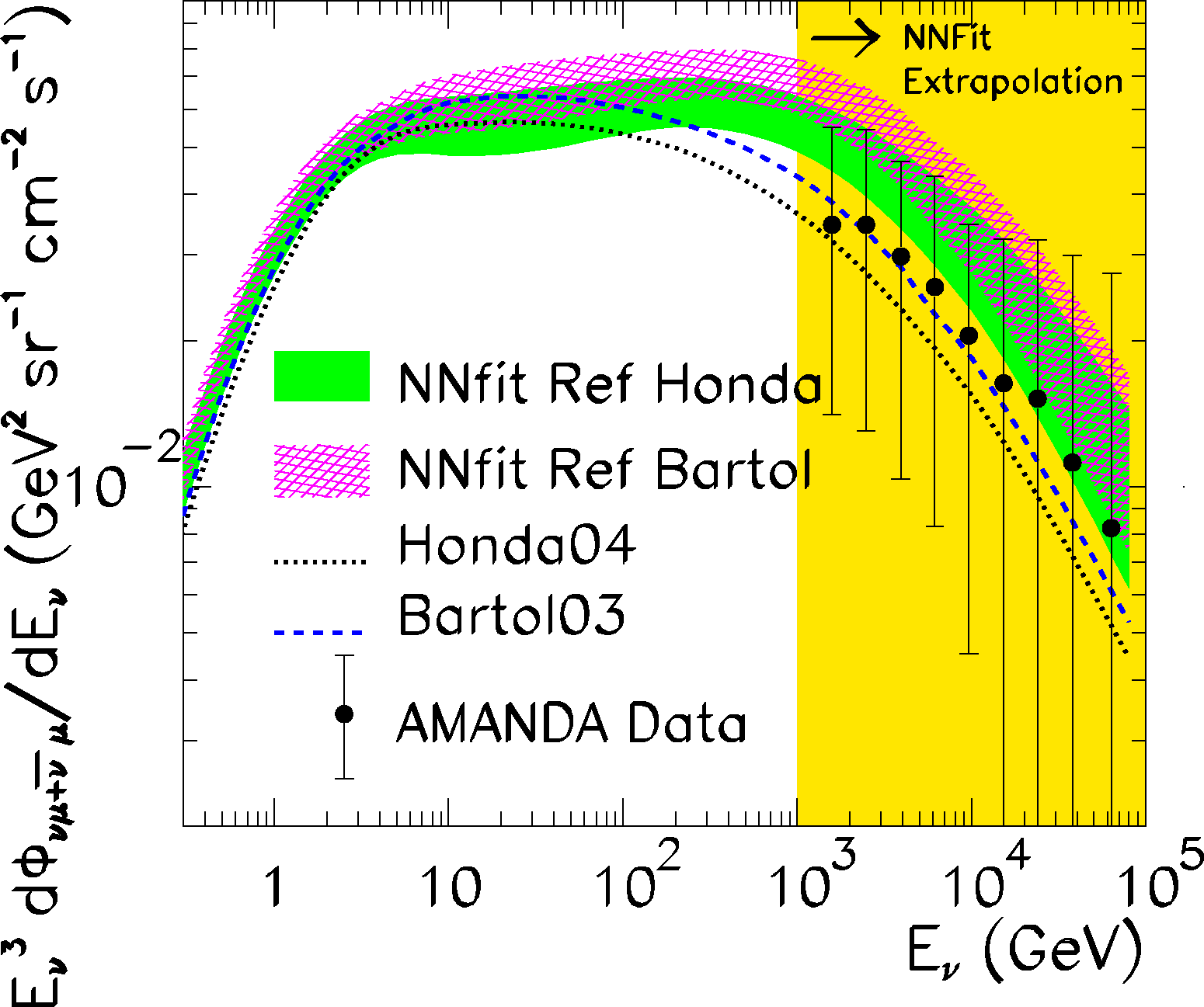}
    \caption{\label{fig:fluxes}%
      Results for the reference fit for the angular averaged muon
      neutrino plus antineutrino flux extrapolated to the high-energy
      region compared to the corresponding data from AMANDA. Also
      shown are the results with different reference fluxes and the
      comparison with the predictions of different flux
      models.\hfill~}
\end{figure}

So far we have discussed the potentialities of atmospheric neutrino
for what concerns the determination of the neutrino parameters. The
main message of the first part of this talk is that atmospheric data
can provide very useful information on the neutrino mass matrix,
despite the very large uncertainties in the neutrino fluxes. However,
it is logically acceptable to invert the strategy, and to regard the
poorly known atmospheric neutrino fluxes as a subject of investigation
themselves. In this section we will therefore assume that the neutrino
parameters have been accurately measured by other experiments, and we
will show that it is possible to extract the the atmospheric neutrino
fluxes directly from the data.\,\cite{GonzalezGarcia:2006ay}

There are several motivations for such direct determination of the
atmospheric neutrino fluxes.  First of all it would provide a
cross-check of the standard flux calculations as well as of the size
of the associated uncertainties (which, being mostly theoretical, are
difficult to quantify). Second, a precise knowledge of the atmospheric
neutrino flux is of importance for high energy neutrino telescopes,
both because they are the main background and they are used for
detector calibration. Finally, such program may quantitatively expand
the physics potential of future atmospheric neutrino experiments.
Technically, however, this program is challenged by the absence of a
generic parametrization of the energy and angular functional
dependence of the fluxes which is valid in all the range of energies
where there is available data. We bypass this problem by using
\emph{artificial neural networks} as unbiased interpolants for the
unknown neutrino fluxes. However, the precision of the available
experimental data is not yet enough to allow for a separate
determination of the energy, zenith angle and flavor dependence of the
atmospheric flux. Consequently in our work we have assumed the zenith
and flavor dependence of the flux to be known with some precision and
extract from the data only its energy dependence. Thus the neural
network flux parametrization will be:
\begin{equation} \label{eq:fluxnn}
    \Phi^\text{fit}_{\alpha,\pm}(E_\nu, c_a, h) =
    F^\text{fit}(E_\nu) \,
    \Phi^\text{ref}_{\alpha,\pm}(E_\nu, c_a, h)
\end{equation}
where $F^\text{fit}(E_\nu)$ is the neural network output when the
input is the neutrino energy $E_\nu$.

In Fig.~\ref{fig:fluxes} we show the results of our fit to the
atmospheric neutrino flux as compared with the computations of the
Honda\,\cite{Honda:2004yz} and Bartol\,\cite{Barr:2004br} groups. The
results of the neural network fit are shown in the form of a $1\sigma$
band, and plotted as a function of the neutrino energy. For comparison
we also show the data from AMANDA.\,\cite{Achterberg:2005fs}
We see from this figure that the flux obtained from the fit is in
reasonable agreement with the theoretical calculations, although the
fit seems to prefer a slightly higher flux at higher energies. This
indicates that until about $E_\nu\sim 1$ TeV we have a good
understanding of the normalization of the fluxes, and that the present
accuracy from Super-Kamiokande neutrino data is comparable with the
theoretical uncertainties from the numerical calculations.
Note also that the results of our alternative fits depends only mildly
on the choice of Honda or Bartol as the reference flux. This suggests
that the present uncertainties on the angular dependence have been
properly estimated, so that the assumed angular dependence has very
little effect on the determination of the energy dependence of the
fluxes. Thus the atmospheric neutrino flux determined with our method
could be used as an alternative to the existing flux calculations.

\section{Conclusions}
\label{sec:conclusions}

In this talk we have discussed the potentialities of atmospheric
neutrino data in the context of future neutrino experiments. We have
shown that despite the large uncertainties in the neutrino fluxes
atmospheric data will still provide useful information on the neutrino
parameters, due to their very broad range in neutrino energy and nadir
angle. In particular, we have proved that the sensitivity obtained by
a combination of atmospheric and long-baseline data is much stronger
than the one achievable by each data set separately. Finally, we have
shown that present atmospheric data can be used to obtain a direct
determination of the atmospheric neutrino fluxes.

\section*{Acknowledgment}

Work supported by MCYT through the Ram\'on y Cajal program, by
CiCYT through the project FPA2006-01105 and by the Comunidad
Aut\'onoma de Madrid through the project P-ESP-00346.


\begin{thebibliography}{9}

\bibitem{GonzalezGarcia:2007ib}
  M.~C.~Gonzalez-Garcia and M.~Maltoni,
  arXiv:0704.1800 [hep-ph].
  
\bibitem{Akhmedov:2006hb}
  E.~Kh.~Akhmedov \EtAl,
  JHEP {\bf 05} (2007) 077
  [arXiv:hep-ph/0612285].

\bibitem{Campagne:2006yx}
  J.~E.~Campagne \EtAl,
  JHEP {\bf 04} (2007) 003
  [arXiv:hep-ph/0603172].

\bibitem{Huber:2005ep}
  P.~Huber \EtAl,
  Phys.\ Rev.\  D {\bf 71} (2005) 053006
  [arXiv:hep-ph/0501037].

\bibitem{GonzalezGarcia:2004cu}
  M.~C.~Gonzalez-Garcia \EtAl,
  Phys.\ Rev.\  D {\bf 70} (2004) 093005
  [arXiv:hep-ph/0408170].

\bibitem{GonzalezGarcia:2006ay}
  M.~C.~Gonzalez-Garcia \EtAl,
  JHEP {\bf 10} (2006) 075
  [arXiv:hep-ph/0607324].

\bibitem{Honda:2004yz}
  M.~Honda \EtAl,
  Phys.\ Rev.\  D {\bf 70} (2004) 043008
  [arXiv:astro-ph/0404457].

\bibitem{Barr:2004br}
  G.~D.~Barr \EtAl,
  Phys.\ Rev.\  D {\bf 70} (2004) 023006
  [arXiv:astro-ph/0403630].

\bibitem{Achterberg:2005fs}
  A.~Achterberg \EtAl\ [IceCube Collaboration],
  arXiv:astro-ph/0509330.

\end{thebibliography}
\end{document}